%% file: paper.tex
\documentclass[journal,a4paper]{IEEEtran}

\usepackage{gnuplottex}
\usepackage{filecontents}
\usepackage{pgfplots, pgfplotstable}
\usepackage{pstricks-add,pst-node,pst-tree}
\usepackage{epsfig}
\usepackage{pst-grad} 
\usepackage{pst-plot} 
\usepackage[space]{grffile} 
\usepackage{etoolbox} 
\makeatletter 
\patchcmd\Gread@eps{\@inputcheck#1 }{\@inputcheck"#1"\relax}{}{}
\makeatother
\usepackage{mathdots}
\usepackage[english]{babel}
\usepackage{relsize}
\usepackage{graphics,graphicx}

\usepackage{url}

\usepackage[breaklinks]{hyperref}
\usepackage{breakurl}

\usepackage{algorithm}
\usepackage{algorithmic}

\usepackage{booktabs}
\usepackage{multirow}
\usepackage{mparhack}
\usepackage{subfigure}
\usepackage{authblk}
\usepackage{mathrsfs}

\usepackage{acronym}

\newcommand{\reQL}[1]{}
\newcommand{\coQL}[1]{}

\usepackage{array}
\usepackage{cite}
\usepackage{eqparbox}
\usepackage{mdwmath}

\usepackage{epsfig}
\usepackage{xcolor}

\usepackage{mathtools,cuted}
\usepackage{bm}
\usepackage{bbold}
\usepackage[all]{xy}
\usepackage{etoolbox}

\usepackage{clrscode3e}
\usepackage{accents}
\usepackage{multicol}

\usepackage{lipsum,color}

\usepackage{tcolorbox}

\usepackage{cite}

\usepackage{pgfplots}
\usepackage{tikz}
\usetikzlibrary{plotmarks}
\usetikzlibrary{patterns}
\usetikzlibrary{shapes.geometric}

\DeclareMathAlphabet{\mathantt}{OT1}{antt}{li}{it}
\DeclareMathAlphabet{\mathpzc}{OT1}{pzc}{m}{it}

\usepackage{accents}

\newtheorem{theorem}{Theorem}

\newcommand\B{\rule[-1.0ex]{0pt}{0pt}}

\DeclareFontFamily{OT1}{pzc}{}
\DeclareFontShape{OT1}{pzc}{m}{it}%
  {<-> s * [1.1] pzcmi7t}{}
\DeclareMathAlphabet{\mathpzc}{OT1}{pzc}%
                     {m}{it}

\def\B{\mathcal{B}}

\def\K{\mathcal{K}}
\def\I{\mathcal{I}}
\def\S{\mathcal{S}}

\DeclareMathOperator{\argmax}{\arg\max}

\renewcommand{\vec}[1]{\mathbf{#1}}

\begin{document}

\makeatletter
\patchcmd{\@maketitle}
  {\addvspace{0.5\baselineskip}\egroup}
  {\addvspace{-1.45\baselineskip}\egroup}
  {}
  {}
\makeatother

\title{Resource Optimization with Flexible Numerology and Frame Structure for Heterogeneous Services}


\author[1]{Lei You}
\author[2]{Qi Liao}
\author[3]{Nikolaos Pappas}
\author[3]{Di Yuan}
\affil[1]{{\small Department of Information Technology, Uppsala University, Sweden}}
\affil[2]{{\small Nokia Bell Labs, Stuttgart, Germany}}
\affil[3]{{\small Department of Science and Technology, Link\"{o}ping University, Sweden}}
\affil[ ]{\texttt{{\scriptsize\{lei.you\}@it.uu.se}}\quad\texttt{{\scriptsize qi.liao@nokia-bell-labs.com}}\quad\texttt{{\scriptsize \{nikolaos.pappas, di.yuan\}@liu.se}}\/}

\tikzset{
    vertex/.style = {
        circle,
        fill            = black,
        outer sep = 2pt,
        inner sep = 1pt,
    }
}

\maketitle

\begin{abstract}
We explore the potential of optimizing resource allocation with
flexible numerology in frequency domain and variable frame structure
in time domain, in presence of services with different types of
requirements. We prove the NP-hardness of the problem, and propose a
scalable optimization algorithm based on linear programming and
Lagrangian duality. Numerical results show significant advantages of
adopting flexibility in both time and frequency domains for capacity
enhancement and meeting the requirements of mission critical services.
\end{abstract}

\section{Introduction}

The \ac{5G} of wireless communications systems is required to support
a large variety of services~\cite{soret2014fundamental}.  A
promising solution for higher resource efficiency while providing
lower latency is the scalable \acp{TTI}~\cite{PedersenVTC2015-fall,
PocoviICCW2017,PedersenVTC2017,liao2016resource,FountoulakisWiOpt2017,2017arXiv171205344A}.
These works fall within the general notion of flexible resource
allocation in the time-frequency domain,
Optimization along the frequency dimension
yields similar structures to problems such as multi-dimensional
\textsc{Knapsack} or weighted \textsc{Matching}
\cite{5165384}. Resource optimization adopting
flexibility in both dimensions regarding frequency and time, named
\textit{2-dimensional (2-D) resource allocation}, poses new
challenges~\cite{4531853,1677815}.   Although flexible resource allocation along
both the time and frequency dimensions is not
new~\cite{4531853,1677815,7829438,7744816,7888960,8263598,7499809},
\textit{from an integer programming point of view, frequency selective
resource allocation with flexible sizes of resource units along both
the frequency and time dimensions}, has not yet been addressed to the
best of our knowledge.

Based on 3GPP release for scalable numerologies and frame
structures~\cite{3GPP36101}, we consider the resulting 2-D resource
allocation problem. We address tractability and propose \textit{an
algorithm with scalability}, utilizing both the \textit{primal} space and \textit{dual}
space of optimization.  We then provide numerical results
for performance assessment.


\section{System Model}
\label{sec:system_model}

Consider a base station and two categories of services. The first
category, denoted by $\K^{(\ell)}$, has strict latency requirement.
For any service $k\in\K^{(\ell)}$, denote its data demand by
$q_k$ (in bits) that has to be met with a latency tolerance of $\tau_k$. Here,
the latency tolerance refers to the time until the data has been fully transmitted
by the scheduler. The parameter can be set to also account for
queueing delay that has taken place, along with the time at the receiver for processing/computing, to
meet the overall deadline of delivery. The second category of
services is denoted by $\K^{(c)}$, for which the target is to maximize
the throughput. For services in $\K^{(c)}$, full buffer is assumed.
Moreover, services in $\K^{(\ell)}$ are prioritized over
those in $\K^{(c)}$. We define $\K=\K^{(\ell)}\cup\K^{(c)}$ the set of all
services.

\begin{figure}[ht!]
\centering
\includegraphics[width=0.67\linewidth]{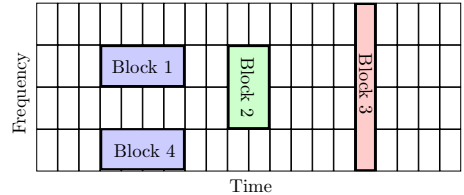}
\caption{An illustration of resource allocation with three types of blocks. A rectangle of the grid is 
a basic unit of the resource.  A service is allocated with one or multiple blocks, and each block can 
be assigned to up to one service. Note that blocks $1$ and $4$ are different blocks with the same shape.}
\label{fig:RU}
\end{figure}

In 5G new radio (NR), a numerology is defined by \ac{SCS} and \ac{CP} length (a.k.a.~the
``guard interval'' between the symbols). The radio frame structure is characterized by number of slots within a frame. A \ac{TTI} can consist of one mini-slot with 1-13 symbols supported, or one slot with 14 symbols (or 12 symbols in case of extended \ac{CP}), or multiple slots if slot aggregation is supported. One resource allocation to a service involves 
a set of adjacent \acp{SCS} and \acp{TTI} in the frequency and time
domain respectively with a configured \ac{CP} length. For simplicity,
hereafter we refer to the resource configuration of numerology and
frame structure as \textit{blocks}, and consider a candidate set $\B$ of {blocks},
see \figurename~\ref{fig:RU}.  For each $b\in\B$, the achieved
throughput on block $b$ if $b$ is assigned to service $k$ ($k\in\K$)
is denoted by $r_{b,k}$. 

Given the channel profile, the transmission power, and the noise
power, $r_{b,k}$ depends on the configuration of block
$b$, including the time span and frequency range (characterized by
\ac{SCS} and \ac{TTI} duration), \ac{CP} length, and symbol
duration. Moreover, this rate shall take into account the effect of guardband.
To compute the achieved throughput per block, we assume a total number of nine multipath channel profiles
\cite[Table B.2.1-4]{3GPP36101}, and we predefine the 
mapping from the configuration parameters to the throughput based on
the model in~\cite{batariere2004cyclic}. This model takes into account the inter-symbol-interference (ISI) depending on CP, and approximates the inter-channel interference (ICI) between the neighboring subbands with the same type of numerology (the ICI between subbands with different types of numerologies is not modeled in this paper due to the high complexity). In addition, we also consider the control overhead as one or more consecutive symbols per TTI. Due to limited space, we
omit the details but provide the tutorial and source code in
\href{http://dx.doi.org/10.21227/ch8e-x385}{IEEE DataPort}~\cite{ch8e-x385-18}.

\section{Problem Formulation and Tractability}
\label{sec:formulation}

Consider the problem of maximizing the total throughput for
$\K^{(c)}$, subject to latency and the demand constraints for
$\K^{(\ell)}$. We use basic unit to refer to the minimum unit of resource 
in the time-frequency domain in the problem formulation.
The set of basic units is denoted by $\I$. We let $a_{b,i}=1$ if block $b$ includes basic unit $i$, otherwise $a_{b,i}=0$. Recall that a block refers to a rectangular shape located at some specific location in the resource grid, see Figure~\ref{fig:RU}. Taking the figure as an example, there are $20 \times 4 = 80$ basic units. Hence $\I = \{1, \dots, 80\}$. Block 1 is of shape $1 \times 4$, and for this block, $a_{1, i}=1$ for four elements of $\I$. Placing this $1 \times 4$ shape at all possible positions of the grid, we obtain all blocks and the corresponding $a$-values for this particular shape. Doing so for all shapes generates the set of all candidate blocks $\B$. As it can be seen from Figure \ref{fig:RU}, for each block, the $a$-values  are fully determined by the position of the block and the numerical indexing of the basic units, and the complexity of doing one mapping equals the size of $\I$, i.e., $|{\I}|$. The complexity for obtaining all the mappings is $O(|{\B}| |{\I}|)$. One possible implementation of performing the mapping is detailed in IEEE DataPort \cite{ch8e-x385-18}. This mapping is done  in the pre-processing stage once (not every TTI), and it is totally decoupled from any service to be scheduled, as it only concerns the two sets $\B$ and $\I$.

The optimization task is to select blocks for each service,
such that the latency and demand requirements are met for $\K^{(\ell)}$, without overlapping among chosen blocks.
We use
optimization variable $x_{b,k}\in\{1,0\}$ to indicate whether block $b$
($b\in\B$) is assigned to the service $k$ ($k\in\K$). A block $b$ is
infeasible for $k$ ($k\in\K^{(\ell)}$), if the ending time of $b$
exceeds $\tau_k$. This is modeled by setting $r_{b,k}=0$. The problem
is formulated below. The two sets of constraints impose the
demand for $\K^{(\ell)}$ and block non-overlapping, respectively.
\begin{subequations}
\begin{alignat}{2}
[\textsc{P0}]\quad &
\max\limits_{\vec{x} \in \{0,1\}} \sum_{b\in\B}\sum_{k\in\K^{(c)}}r_{b,k}x_{b,k}  \\
\text{s.t.} \quad & \sum_{b\in\B} r_{b,k}x_{b,k} \geq q_{k},~k\in\K^{(\ell)} \label{eq:P0-demand} \\
				  & \sum_{k\in\K}\sum_{b\in\B} a_{b,i}x_{b,k}\leq 1,~i\in\I \label{eq:P0-RU}
\end{alignat}
\label{eq:P0}
\end{subequations}
\begin{theorem}
\textsc{P0} is $\mathcal{NP}$-hard.
\label{thm:np-hard}
\end{theorem}
\begin{IEEEproof}
We construct a polynomial-time reduction from the \textsc{Partition
Problem (PP)} for a set of integers $\{d_1, \dots, d_n\}$. The task is
to determine whether or not there is a partition such that the two
subsets have equal sum of ${\sum_{i=1}^n d_i}/{2}$, where the
numerator is assumed to be even. We define a single TTI size and multiple
blocks that take the shape of one basic unit.  There are two services,
denoted by $k^\ell$ and $k^c$, in $\K^{(\ell)}$ and $\K^{(c)}$,
respectively. The latency parameter of $k^\ell$ equals that of the TTI
size, and the demand equals ${\sum_{i=1}^n d_i}/{2}$.
Moreover, $r_{b,1} = r_{b,2} = d_b, b =1, \dots,n$. By
construction, \eqref{eq:P0-RU} has no effect. Next, one can observe that
partitioning the $n$ basic units into two subsets, each providing
a total throughput of ${\sum_{i=1}^n d_i}/{2}$, is equivalent
to a feasible solution to PP. In addition, this can occur
if and only if the objective function defined for $k^c$ reaches ${\sum_{i=1}^n d_i}/{2}$.
Hence the conclusion.
\end{IEEEproof}

\section{Problem Solving}
\label{sec:solving}

We propose a sub-optimal but low-complexity algorithm, 
consisting in performing assignment of blocks to services, based on
utility values generated from \textit{linear programming (LP)
relaxation and the Lagrangian dual (LD)}. 

\subsection{\textcolor[rgb]{0,0,0}{Block Assignment}}
\label{subsec:block-assignment}

We denote by matrix $\vec{u}$ of size $|\B|\times |\K|$ the
\textit{utility matrix} for all pairs of blocks and services. 
An element $u_{b,k}$ represents the utility of a block-service pair
$(b,k)$ ($b\in\B$ and $k\in\K$). Block assignments for $\K^{(\ell)}$ and
$\K^{(c)}$ are treated separately. The former is performed first
because of the latency requirement. 

\begin{algorithm}
\renewcommand{\algorithmicrequire}{\textbf{Input:}}
\renewcommand{\algorithmicensure}{\textbf{Output: }}
\begin{algorithmic}[1]
\caption{\textcolor[rgb]{0,0,0}{\textsc{BA($\S$,$\vec{u}$)}. 
{The input consists of a block-service assignment $\S$ which may be empty, and 
a utility matrix $\vec{u}$. Set $\S$ is augmented and then returned
by the algorithm. We remark that $b$ overlaps with $b'$ if and only if $\exists i\in\I~a_{b,i}+a_{b',i}>1$.} \label{alg:greedy}	
}}

\REPEAT \label{alg:repeat1}
\STATE Remove from $\B$ the blocks in $\S$ and those overlapping with the blocks in $\S$.   \label{alg:initial_set}
\STATE  \textcolor[rgb]{0,0,0}{$(b',k')\leftarrow\argmax_{b\in\B,k\in\K^{(\ell)}}u_{b,k}$, $\S\leftarrow\S\!\cup\!\{(b',k')\}$.}\label{alg:finding}
\STATE \textbf{if} $q_{k'}$ is met: $\K^{(\ell)}\leftarrow \K^{(\ell)}\backslash\{k'\}$.  \label{alg:demand_met}
\UNTIL{\textcolor[rgb]{0,0,0}{$\K^{(\ell)}=\phi$} \textbf{or} \textcolor[rgb]{0,0,0}{$\B=\phi$}}\label{alg:until1}
\STATE \textbf{if} $\K^{(\ell)}\neq\phi$:
\STATE \quad The demand of the users left in $\K^{(\ell)}$ cannot be met.
\REPEAT \label{alg:repeat2}
\STATE \textcolor[rgb]{0,0,0}{Lines~\ref{alg:initial_set}--\ref{alg:finding} with the notation $\K^{(\ell)}$ replaced by $\K^{(c)}$.}
\UNTIL{\textcolor[rgb]{0,0,0}{$\B=\phi$}} \label{alg:until2}
\end{algorithmic}
\end{algorithm}

The 
operations in Lines~\ref{alg:initial_set} and \ref{alg:demand_met}
cost $O(1)$ by hash-map implementations. 
By sorting the utilities in
advance (which costs
$O\left(|\B||\K^{(\ell)}|\log\left(|\B||\K^{(\ell)}|\right)\right)$),
along with inspecting the other algorithm operations,
one can conclude that the overall complexity
is of $O\left(|\B||\K|\log\left(|\B||\K|\right)\right)$.

\subsection{Utility Estimation by LP Relaxation}
\label{subsec:primal}

One way to compute the utility matrix $\vec{u}$ is to
solve the LP relaxation of \textsc{P0} and to use the LP optimum
$\vec{x}_{\text{LP}}$.
\begin{subequations}
\begin{alignat}{2}
[\textsc{P0-LP}]\quad &
\max_{0 \leq \vec{x} \leq 1}\sum_{b\in\B}\sum_{k\in\K^{(c)}}r_{b,k}x_{b,k}~ 
\text{s.t.}~ \eqref{eq:P0-demand},~\eqref{eq:P0-RU} 
\end{alignat}
\label{eq:P0-LP}
\end{subequations}
\!\!\!\!\!
We denote by $\vec{u}_{\text{LP}}=\vec{x}_{\text{LP}}$ the LP-based
utility. Also, $\vec{x}_{\text{LP}}$ can be used for initialization:
$\S=\{(b,k):u_{\text{LP},b,k}\geq \rho, b\in\B, k\in\K\}$ with $\rho$
being a threshold.

\subsection{Utility Estimation by LD}
\label{subsec:dual}

By relaxing the constraints~\eqref{eq:P0-RU} of \textsc{P0} with Lagrangian multiplier $\lambda_i$ ($i\in\I$), the Lagrangian is defined as follows:
\[
L(\vec{x},\bm{\lambda}) =\sum_{b\in\B}\sum_{k\in\K^{(c)}}r_{b,k}x_{b,k}  
+ \sum_{i\in\I}\lambda_i \left(1-\sum_{b\in\B}\sum_{k\in\K}a_{b,i}x_{b,k}\right) 
\]
The LD function is defined in~\eqref{eq:P1}.
\begin{equation}
[\textsc{P1}]\quad g(\bm{\lambda}) =\max\limits_{\vec{x} \in \{0,1\}} L(\vec{x},\bm{\lambda})~\text{s.t.}~\eqref{eq:P0-demand}
\label{eq:P1}
\end{equation}
\!\!\!
Accordingly, we have the LD problem:
\begin{equation}
[\textsc{P0-LD}]\quad \min_{\bm{\lambda}\geq\vec{0}} g(\bm{\lambda})	.
\end{equation}
We define 
$\alpha_{b}=\sum_{i\in\I}\lambda_i a_{b,i}$. 
Problem \textsc{P1} decomposes for $\K^{(c)}$ and $\K^{(\ell)}$: 
The constraints~\eqref{eq:P0-demand} are only for~$\K^{(\ell)}$. Therefore, solving~\textsc{P1} amounts to solving the two problems \textsc{P2} and \textsc{P3} for $\K^{(c)}$ and $\K^{(\ell)}$ respectively, shown below.
\begin{subequations}
\begin{alignat}{2}
[\textsc{P2}]\quad &
\max\limits_{\vec{x}}  \sum_{b\in\B}\sum_{k\in\K^{(c)}}(r_{b,k}-\alpha_{b})x_{b,k}  \label{eq:P2-obj} \\ 
\text{s.t.} \quad & \sum_{k\in\K^{(c)}} x_{b,k}\leq 1,~b\in\B \label{eq:P2-b} \\
				  & x_{b,k}\in\{0,1\},~b\in\B,~k\in\K^{(c)} \label{eq:P2-x}
\end{alignat}
\label{eq:P2}
\end{subequations}
\begin{subequations}
\begin{alignat}{2}
[\textsc{P3}]\quad &
\min\limits_{\vec{x} \in \{0,1\}} \sum_{b\in\B}\sum_{k\in\K^{(\ell)}}\alpha_{b}x_{b,k}  \\ 
\text{s.t.} \quad & \sum_{b\in\B} r_{b,k}x_{b,k} \geq q_{k},~k\in\K^{(\ell)} \label{eq:P3-demand}
\end{alignat}
\label{eq:P3}
\end{subequations}

Note constraints~\eqref{eq:P2-b} are not present in \textsc{P0},
though these are implied by \eqref{eq:P0-RU} for services in
$\K^{(c)}$. 
Computing the optimum of \textsc{P2} is straightforward. Each block $b$ is allocated to the service $\argmax_{k}{r_{b,k}-\alpha_b}$ with $r_{b,k}-\alpha_b>0$. 

Problem \textsc{P3} further decomposes to $\left|\K^{(\ell)}\right|$
problems \textsc{P3[$k$]} ($k\in\K^{(\ell)}$), each with objective
$\max\sum_{b\in\B}\alpha_{b}x_{b,k}$, constraints
$\sum_{b\in\B}r_{b,k}x_{b,k}\geq q_{k}$, and binary variables
$x_{b,k}$ ($b\in\B$).
\begin{subequations}
\begin{alignat}{2}
[\textsc{P3[$k$]}]\quad &
\min\limits_{\vec{x} \in \{0,1\}} \sum_{b\in\B}\alpha_{b}x_{b,k}  \\ 
\text{s.t.} \quad & \sum_{b\in\B} r_{b,k}x_{b,k} \geq q_k \label{eq:P3-k-demand} 
\end{alignat}
\label{eq:P3-k}
\end{subequations}
Each \textsc{P3[$k$]} can be reformulated as a \textsc{Knapsack
Problem}, and \textit{optimally}
solved by dynamic programming.

The dual problem \textsc{P0-LD} can be
solved using a \textit{sub-gradient method}~\cite{Sen1986}.  Denote by
$\vec{x}^{(h)}_{\text{LD}}$ the LD solution in the $h_{\text{th}}$
iteration of the sub-gradient method. We let
$\vec{u}_{\text{LD}}=\sum_{h}\vec{x}^{(h)}_{\text{LD}}$ to be the
LD-based utility.

\subsection{\textcolor[rgb]{0,0,0}{Algorithm Implementation}}
\label{subsec:alg}

In addition to \textsc{BA($\S$,$\vec{u}_{\text{LP}}$)} and \textsc{BA($\S$,$\vec{u}_{\text{LD}}$)}, 
we consider algorithm ``LP+LD'' that returns the best solution of
\textsc{BA($\S$,$\vec{u}_{\text{LP}}$)} and \textsc{BA($\S$,$\vec{u}_{\text{LD}}$)}.
We remark that \textsc{BA($\S$,$\vec{u}_{\text{LD}}$)} is quite
flexible in terms of computational effort, as one can use
accumulated $\vec{u}_{\text{LD}}$ before full convergence.
Overall, the algorithm scales well. Moreover, if necessary,
the service sets can be decomposed into subsets, and the algorithm can
be applies to one subset at a time to further reduce complexity.

\section{Numerical Results}
\label{sec:numerical}

The use of flexible numerology is expected to outperform fixed
numerology. The purpose of performance evaluation is to examine the
amount of improvement, which is of significance in particular as the
control channel overhead for supporting the flexible structure is
accounted for. The result also tell how well the proposed
algorithm is suited for the flexible structure.

Comparing to LTE that applies a fixed SCS of 15 kHz and TTI of $1.0$ ms, we consider four shapes, Shape 1, Shape 2, Shape 3, and Shape 4, with SCS being $15$ kHz, $30$ kHz, $60$ kHz, and $60$ kHz, CP $4.7$ $\mu$s, $2.3$ $\mu$s $1.2$ $\mu$s, and $4.17$ $\mu$s, and the number of symbols $7$, $7$, $7$, and $6$, respectively. The TTI durations of the four shapes are $0.5$ ms, $0.25$ ms, $0.125$ ms, and $0.125$ ms, respectively.
The numerologies ($\Delta f = 2^{\mu}\times 15$ kHz, $\mu=0,1,2\ldots$) originate from Release~15 \cite[Table
4.2-1]{3GPP38211}. Note that Release~15 also specifies subcarrier
spacing up to $240$~kHz. However, by \cite[Table I]{8329620}, a TTI of
$0.125$ ms meets all the worst-case transmission latencies for the
listed 5G ultra-reliable low-latency communication configurations. 


Parameter settings
are given in \tablename~\ref{tab:sim}. We test our algorithm for a set
of candidate thresholds $\rho\in\{0.05,0.5,\ldots,0.95\}$ among which
the one achieves the best objective is selected. The maximum sub-gradient
iterations is set to $200$. While calculating the block rates, the
impact on capacity due to guardband is included by following the model
in \cite{8263598}. The rate reduction due to control overhead follows that in
\cite{MiFa17}, where two symbols per TTI constitute the overhead.
We emphasize on accurate assessment in terms
of optimality, that is, how much does the proposed algorithm perform
with respect to global optimum. We use the global optimum obtained by
solving the integer programming problem~\eqref{eq:P0} via a solver.
This is not a scalable method. The purpose here is for benchmarking,
to demonstrate that our low-complexity algorithm has little loss in
optimality.  The number of users as well as the bandwidth is chosen
such that the global optimum can be obtained with reasonable amount of
computing effort.  Similarly, in view of the computational effort of
obtaining global optimum for benchmarking, we do not include all TTI
sizes that are permitted by 5G NR \cite{3GPP38211}.

\begin{table}[!h]
\centering
\caption{Simulation Parameters.}
\begin{tabular}{ll}
\toprule
\textbf{Parameter} & \textbf{Value} \\
Number of users & 10 with $|\K^{(\ell)}|=|\K^{(c)}|=5$ \\
Time-frequency domain & $2$ ms and $2$ MHz\\
SNR range & $[5,30]$ (dB) \\
Demand $q$ & $\{16,32,64,128,256,512\}$ (kbps) \\
Latency tolerance $\tau$ & $\{0.25, 0.5, 1.0, 1.5, 2\}$ (ms) \\
Threshold (Section~\ref{subsec:primal}) & \textcolor[rgb]{0,0,0}{$\rho\in\{0.05,0.1,\ldots,0.95\}$} \\
\bottomrule
\end{tabular}
\label{tab:sim}	
\end{table}

\figurename~\ref{fig:bitrate} 
shows the average bit rate of services in $\K^{(c)}$ with respect to the
latency tolerance $\tau$ of $\K^{(\ell)}$. For the non-flexible
structures, Shape 1, Shape 2, and Shape 3 are used separately. Each of these
structures is referred to in the format of ``TTI-SCS''
(e.g. $0.25$ms-$30$kHz means a shape of a fixed TTI of $0.25$ ms and a
fixed SCS of $30$ kHz).  

The flexible structure significantly outperforms the non-flexible
ones. The system tends to benefit more from flexible structure when
the latency tolerance becomes more stringent. Note that our 
algorithm with flexible structure is near-optimal.
Among the three non-flexible schemes, $0.25$ms-$30$kHz outperforms
the other two. This result is related to that, in the optimization
problem, the throughput of the services in $\K^{(c)}$, is subject to
latency constraints of services in $\K^{(\ell)}$.  For resource
allocation, blocks of $0.5$ms-$15$kHz and $0.125$ms-$60$kHz have low
flexibility in the time and frequency domains, respectively.  The
former does not have many choices in meeting the latency requirements
of $\K^{(\ell)}$, leading to poor throughput for $\K^{(c)}$. The
latter is not efficient on the frequency domain (due to frequency
selectivity), though this inefficiency is mitigated when the latency
tolerance is high, as more choices become available in the time
domain.  Blocks of $0.25$ms-$30$kHz strikes a balance between short
TTI size (to meet latency-constrained services) and flexibility in the
frequency domain.

Without showing by the figures, we remark that the
problem feasibility of the three non-flexible schemes is very
sensitive to the latency tolerance. This issue is alleviated by the
flexible structure. In comparison to the related
work \cite{7417854} considering advantages of flexible numerology,
 our results emphasize the significance
of block-service assignment optimization.

\pgfplotsset{compat=1.11,
        /pgfplots/ybar legend/.style={
        /pgfplots/legend image code/.code={%
        \draw[##1,/tikz/.cd,bar width=3pt,yshift=-0.2em,bar shift=0pt]
                plot coordinates {(0cm,0.8em)};},
},
}
\begin{figure}[h!]
\centering
\begin{tikzpicture}
\begin{axis}[
	label style = {font=\fontsize{9pt}{10pt}\selectfont},
    ybar,
    legend style={ font=\fontsize{7.8pt}{10pt}\selectfont,
      anchor=north,legend columns=3},
      legend pos = north west,
	enlarge x limits=0.15,
	ymin = -10,
	ymax = 750,
	xlabel={Latency tolerance $\tau$ of $\K^{(\ell)}$ (ms)},
	ylabel={Bit rate per user in $\K^{(c)}$ (kbps)},
    symbolic x coords={0.25, 0.5, 1.0, 1.5, 2.0},
    xtick=data,
	minor x tick num=0,
	minor y tick num=4,
	major tick length= 0.15cm,
	minor tick length=0.075cm,
	tick style={semithick,color=black},
	height=0.667\linewidth,
	width=\linewidth,
    bar width = 0.15cm,
    legend entries = {0.125ms-60kHz, 0.5ms-15kHz, 0.25ms-30kHz, Flexible (LP+LD), Flexible (Optimal)},
    xtick style={draw=none},
    ]
    
\addplot  coordinates {(0.25, 0) (0.5, 0) (1.0, 0)  (1.5, 193.83) (2.0, 193.83) };
\addplot  coordinates {                (0.5, 21.8825) (1.0, 21.8825)  (1.5, 21.8825) (2.0, 21.8825) };
\addplot  coordinates {(0.25, 59.3841) (0.5, 59.3841) (1.0, 64.8245)  (1.5, 234.24) (2.0, 234.24) };
\addplot  coordinates {(0.25, 439.262) (0.5, 453.316) (1.0, 469.637)  (1.5, 470.19) (2.0, 472.907) };
\addplot  coordinates {(0.25, 446.992) (0.5, 467.452) (1.0, 489.134)  (1.5, 502.523) (2.0, 507.62) };

\end{axis}
\end{tikzpicture}
\caption{The figure shows bit rate of $\K^{(c)}$ with respect to latency tolerance $\tau$ of $\K^{(\ell)}$. The bit rate demand of $\K^{(\ell)}$ equals $128$ kbps. All the three cases of non-flexible structures are solved to optimality. When $\tau=0.25$, using the block 0.5ms-15kHz results in infeasibility.}
\label{fig:bitrate}
\end{figure}
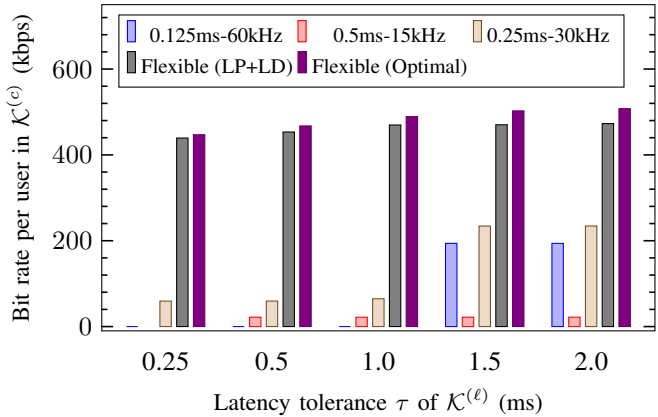

\figurename~\ref{fig:optimality} shows the optimality gaps
as function of the demand of $\K^{(\ell)}$. Here we also include
\textsc{BA($\phi$,$\vec{r}$)}, which uses the throughput of each
block-service pair as the utility.
One can observe that in general the
optimality gap increases with the user demand. Meanwhile, searching in
the dual space for computing block-service utilities in most cases
leads to significantly better results than considering the LP
relaxation.  With high user demand, \textsc{BA($\phi$,$\vec{r}$)} and
\textsc{BA($\S$,$\vec{u}_{\text{LP}}$)} are clearly inferior to the
others. Basically, using LD for utility estimation significantly
reduces the optimality gap. In addition, combining 
\textsc{BA($\S$,$\vec{u}_{\text{LP}}$)},
\textsc{BA($\S$,$\vec{u}_{\text{LD}}$)} leads to further optimality
gap reduction, indicating that LP and LD are complementary to each
other.  Overall, the gap of LP+LD is below
$10$\%.

\pgfplotsset{compat=1.11,
        /pgfplots/ybar legend/.style={
        /pgfplots/legend image code/.code={%
        \draw[##1,/tikz/.cd,bar width=3pt,yshift=-0.2em,bar shift=0pt]
                plot coordinates {(0cm,0.8em)};},
},
}
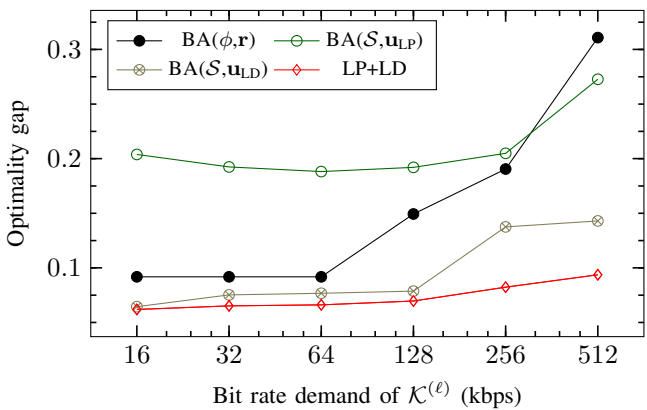
\begin{figure}[h!] 
\centering 
\begin{tikzpicture}
\begin{axis}[
	xmode=log,
	log ticks with fixed point,
	xlabel={Bit rate demand of $\K^{(\ell)}$ (kbps)},
	ylabel={Optimality gap},
	label style = {font=\fontsize{9pt}{10pt}\selectfont},
    legend style={ font=\fontsize{7.8pt}{10pt}\selectfont,
      anchor=north,legend columns=2},
      legend pos = north west,
minor x tick num=4,
minor y tick num=4,
major tick length=0.15cm,
minor tick length=0.075cm,
tick style={semithick,color=black},
	height=0.667\linewidth,
	width=\linewidth,
        scaled y ticks=false,
		xtick = {16,32,64,128,256,512},
		ytick = {0,0.1,...,0.6},
		minor ytick={0,0.025,...,0.6},
		minor xtick={20, 24, 28, 32, 40, 48, 56, 64, 80, 96, 112, 128, 160, 192, 224, 256, 320, 384, 448}
]


\addplot [mark=*, color=black] coordinates {
	(16, 0.0917778548305)
	(32, 0.0917778548305)
	(64, 0.0917778548305)
	(128, 0.149298808848)
	(256, 0.190471082817)
	(512, 0.3108237643)
};

\addplot [ mark=o, color=black!60!green] coordinates {
	(16, 0.20382959834)
	(32, 0.192399580868)
	(64, 0.188153612261)
	(128, 0.192002354004)
	(256, 0.204849702857)
	(512, 0.2726693878)
};

\addplot [mark= otimes, color=black!60!yellow] coordinates {
	(16, 0.0644509718293)
	(32, 0.0752375442755)
	(64, 0.0767109803676)
	(128, 0.0786719629306)
	(256, 0.137497871723)
	(512, 0.142954523237)
};

\addplot [ mark= diamond, color=red] coordinates{
	(16, 0.0618985732275)
	(32, 0.0651771452258)
	(64, 0.0660966)
	(128, 0.0696226593526)
	(256, 0.0822999)
	(512, 0.093656253533)
};

\addplot [ mark= diamond, color=red] coordinates{
	(16, 0.0618985732275)
	(32, 0.0651771452258)
	(64, 0.0660966)
	(128, 0.0696226593526)
	(256, 0.0822999)
	(512, 0.093656253533)
};
\legend{ {\textsc{BA}($\phi$,$\vec{r}$)}, {\textsc{BA}($\S$,$\vec{u}_{\text{LP}}$)}, {\textsc{BA}($\S$,$\vec{u}_{\text{LD}}$)}, LP+LD}
\end{axis}
\end{tikzpicture} 
\caption{The figure shows the optimality gaps of the solutions obtained. For instance, ``0.1'' in the y-axis in \figurename~\ref{fig:optimality} means that the relative deviation to the optimum is 10\% on average. }
\label{fig:optimality}
\end{figure}

\section{Conclusion}
\label{sec:conclusion}

We suggest that combining a flexible numerology and frame
structure serves as a promising option for
spectral efficiency. Utilizing LP and LD 
enables efficient problem solving.

\section*{Acknowledgement}
This work has been partially supported by European Union H2020 MSCA projects ACT5G (643002) and DECADE (645705), and the Center for Industrial Information Technology (CENIIT). The work of the first author was partly accomplished while he was at Link{\"o}ping University, Sweden.

\input{acronyms}

\bibliographystyle{IEEEtran}
\bibliography{ref}

\end{document}

%% file: acronyms.tex
\acrodef{3GPP}{3rd generation partnership project}
\acrodef{5G}{fifth generation}
 \acrodef{ABS}{almost blank subframe}
    \acrodef{BS}{base station}
    \acrodef{CDF}{cumulative distribution function}
    \acrodef{CSI}{channel state information}
    \acrodef{CQI}{channel quality indicator}
		\acrodef{CP}{cyclic prefix}
\acrodef{DL}{downlink}
    \acrodef{DUDe}{downlink and uplink decoupling}
\acrodef{eICIC}{enhanced intercell interference coordination}
\acrodef{ESD}{energy spectral density}
\acrodef{FDD}{frequency division duplex}
    \acrodef{FDMA}{frequency division multiple access}
   \acrodef{GP}{Gaussian process}
    \acrodef{GPS}{global positioning system}
\acrodef{HetNet}{heterogeneous network}
    \acrodef{ICI}{inter-cell interference}
		\acrodef{IMI}{inter-mode interference}
		\acrodef{ISI}{inter-symbol interference}
\acrodef{LTE}{long term evolution}

\acrodef{MAC}{media access control}
\acrodef{MRU}{minimum resource unit}
\acrodef{NR}{new radio}

   \acrodef{OFDM}{orthogonal frequency division multiplexing}
    \acrodef{PDF}{probability density function}
    \acrodef{PHY}{physical layer}
		\acrodef{PSD}{power spectral density}
    \acrodef{PRB}{physical resource block}
		\acrodef{RU}{resource unit}
   \acrodef{QoE}{quality of experience}
    \acrodef{QoS}{quality of service}
    \acrodef{RAN}{radio access network}
		\acrodef{RBS}{removal of bottleneck services}
		\acrodef{RMDI}{resource muting for dominant interferer}
    \acrodef{RRM}{radio resource management}
		\acrodef{RU}{resource unit}
		\acrodef{RX}{receiver}
 \acrodef{SAFP}{successive approximation of fixed point}
    \acrodef{SDN}{software defined network}
    \acrodef{SNR}{signal-to-noise ratio}
    \acrodef{SINR}{signal-to-interference-plus-noise ratio}
\acrodef{SIR}{signal-to-interference ratio}
\acrodef{SIF}{standard interference function}
    \acrodef{SVM}{support vector machine}
		\acrodef{SCS}{sub-carrier spacing}
    \acrodef{TCP}{transmission control protocol}
		\acrodef{TDD}{time division duplex}
    \acrodef{TDMA}{time division multiple access}
		\acrodef{TTI}{transmission time interval}
		\acrodef{TX}{transmitter}
		\acrodef{UE}{user equipment}
		\acrodef{UL}{uplink}
    \acrodef{WLAN}{wireless local area network}